\documentclass[twoside]{dis04}
\def\runauthor{Hannes Jung}
\def\shorttitle{Un-integrated Parton Density Functions  in  CCFM}
\newcommand{\alphasb}{\bar{\alpha}_s}
\newcommand{\Pmax}{\bar{q}}
\newcommand{\kt}{k_{t}}
\newcommand{\pt}{p_{t}}
\newcommand{\ktp}{k_{t}^{\prime}}

\newcommand{\qb}{\ensuremath{\bar{q}} }
\newcommand{\gap}{\stackrel{>}{\sim}}

\def\prp{\perp}

\begin{document}

\title{Un-integrated PDFs in CCFM }

\author{Hannes Jung}

\address{DESY\\ 
Notkestr.85, 22603 Hamburg, FRG\\
E-mail: hannes.jung@desy.de }

\maketitle

\abstracts{The un-integrated parton distribution functions (uPDFs) obtained from
a CCFM evolution are studied
in terms of the intrinsic transverse momentum distribution at low scales.
The uPDFs are studied for variations of the renormalization and factorization
scales.
}
\section{Introduction}
Un-integrated parton distributions are best suited to study the details of the
perturbative as well as the non-perturbative QCD evolution. 
The general form of
the integral equation for the parton evolution is:
\begin{equation}
x{\cal A} (x,\kt,\Pmax ) = x{\cal A}_0 (x,\kt,\Pmax ) + \int \frac{dz }{z} 
\int \frac{d q^2}{ q^{2}} \Theta (\mbox{order})\Delta_s 
\tilde{P}(z,q,\kt) x{\cal A}\left(\frac{x}{z},\ktp,q\right) 
\label{integral} 
\end{equation}  
with $\Pmax$ being
the evolution (factorization) scale. The ordering condition of the evolution  
is specified by $\Theta(\mbox{order})$ which in the case of CCFM is given by
$ \Theta(\mbox{order})=\Theta(\Pmax - zq)$. 
The first term of the rhs of eq.(\ref{integral}) gives the 
contribution of non-resolvable branchings 
between the starting scale $Q_0$ and the factorization scale 
$\Pmax$ and is given by:
\begin{equation}
x{\cal A}_0 (x,\kt,\Pmax ) = x{\cal A}_0 (x,\kt,Q_0) \Delta_s(\Pmax ,Q_0)
\end{equation} 
where the Sudakov form factor 
$\Delta_s(q_2,q_1)$ describes the probability of no radiation between 
$Q_0$ and $\Pmax$. 
The second term of the rhs of eq.(\ref{integral}) describes the details of
the QCD evolution, 
expressed by the convolution of the splitting function $\tilde{P}$ with the
parton density and the Sudakov form factor $\Delta_s$.
\section{The starting distribution \boldmath$x {\cal A}_0(x,\kt,Q_0)$}
The non-resolvable branching contribution is directly related to the initial (starting)
distribution ${\cal A}_0 (x,\kt,Q_0)$, and especially to the intrinsic $\kt$
distribution. Often a Gaussian
type distribution with width $k_0$ is assumed and  ${\cal A}_0$ can be 
parameterized as:
\begin{equation}
x {\cal A}_0(x,\kt,Q_0) = N 
 x^{p_0} (1-x)^{p_1} \cdot \exp{\left(-\kt^2/k_0^2\right)}
\label{eq1}
\end{equation}
where $N$ is a normalization constant, and $p_0, p_1$ being parameters to be
determined experimentally.
\par
In the following sections the dependence of the choice of $p_0, p_1$ and 
$k_0$ will be investigated.  The CCFM evolution equation is applied to the 
starting distribution  $x {\cal A}_0(x,\kt,Q_0)$ and then convoluted with the off-shell
matrix elements. 
The simple gluon splitting function is 
applied~\cite{jung_salam_2000,jung-dis03}, 
\begin{equation}
P_{gg}(z,\qb,\kt )= \frac{\alphasb(q^2(1-z)^2)}{1-z} + 
\frac{\alphasb(\kt^2)}{z} \Delta_{ns}(z,q^2,\kt)
\label{Pgg}
\end{equation}
with $q = \pt/(1-z)$ and 
the non-Sudakov form factor $\Delta_{ns}$ is defined by:
\begin{eqnarray}
\log\Delta_{ns}(z_i,q^2,\kt)& = & -\alphasb
                  \int_{z}^1 \frac{dz'}{z'} 
                        \int \frac{d q^2}{q^2}  \cdot 
                  \Theta(\kt-q)\Theta(q-z'q)
               \label{non_sudakov_int}       
\end{eqnarray}
The collinear cutoff $Q_g= 1.3$~GeV regulates the region of  $z \to 1$, 
applied both to the real emissions as well as in the Sudakov form factor 
for the virtual corrections, which is equivalent to using the {\it plus}
prescription of the splitting function $P_+$.
\par
Due to the angular ordering in CCFM 
a kind of random walk in the propagator gluon 
$\kt$ can be performed, and values of small $\kt$ can be reached. A cut 
$\kt^{cut}$ is introduced to avoid this region.
\par
The CCFM evolution equations have been solved 
numerically~\cite{jung_salam_2000} using a Monte Carlo method. 
In Tab.~\ref{pdfsets} the fit results for the 
un-integrated gluon densities are summarized. 
The input parameters  are varied such
that after convolution with the off-shell matrix elements the results
agree best with the measured structure function $F_2(x,Q^2)$ 
in the range $x < 5 \cdot 10^{-3}$ and  $Q^2 > 4.5$~GeV$^2$ 
as measured at H1~\cite{H1_F2_1996,H1_F2_2001} and  
ZEUS~\cite{ZEUS_F2_1996,ZEUS_F2_2001}. The starting distribution is given by
eq.(\ref{eq1}), where the parameter $p_1=4$ is kept fixed, but $p_0$ and the
width of the Gaussian $k_0$ was adjusted to give the smallest $\chi^2/points$.
The results are given in Tab.~\ref{pdfsets} and labeled as {\it set A0}.
\par
Equally good fits can be obtained using different values for the soft cut
$\kt^{cut}$ and a different value for the width of the intrinsic $\kt$
distribution {\it set B0}. 
Motivated by the success of the saturation model~\cite{GBW}, the
parameterization of GBW can also serve as a starting distribution: 
\begin{equation}
x {\cal A}_0(x,\kt,Q_0) = \frac{3 \sigma_0}{4\pi^2\alpha_s} R_0^2(x) \kt^2 
\exp\left(-R_0^2(x) \kt^2 \right)
\end{equation} 
with the parameters given in~\cite{GBW}. As seen from Tab.~\ref{pdfsets} a
reasonable description of the data can be achieved {\it set GBW}.
\par
The renormalization scale dependence of the final 
cross section can be estimated by changing the scale used in $\alpha_s$ in the
off-shell matrix element. Since here we are using the LO $\alpha_s$ matrix
elements, any scale variation will change the cross section. In order to obtain
a reasonable result, the uPDF was fitted to describe $F_2$  by varying the scale
$\mu_r$. The {\it set A-} ({\it set B-}) correspond to a scale $\mu_r = 0.5 \pt$
and the {\it set A+} ({\it set B+}) correspond to a scale $\mu_r = 2 \pt$. 
 by: $0.5\cdot \mu_r$ and $2\cdot \mu_r$. The resulting $\chi^2/points$
is also shown in Tab.~\ref{pdfsets}. However changing the scale $\mu_r$ and 
$\mu_f$ simultaneously with the same factor results in a much
larger $\chi^2$, which is understandable, since the length of the evolution
ladder is changed by $\mu_f$.
\par 
The CCFM  evolution is performed in an angular ordered region of phase space and
the factorization scale is related to the maximum allowed
angle for any emission. In CCFM the scale $\mu_f$ (coming from the maximum angle)
can be written as:
\begin{equation}
\mu_f  = \sqrt{\hat{s} + Q_{\prp}^2}
\label{evolution-scale}
\end{equation} 
with $\hat{s}$ being the invariant mass of the $q\bar{q}$ subsystem, and 
$Q_{\prp}$ its transverse momentum. In this definition, the scale $\mu_f$ is
related to the quark pair. 
\par
In order to investigate the uncertainties coming from the specific choice of the
evolution scale given in eq.(\ref{evolution-scale}),  
another definition is applied, relating the factorization scale only to the 
quark (or anti-quark):
$\mu_f  = \frac{\pt}{1-z}$
with $\pt$ being the transverse momentum of the quark(anti-quark) and 
$z = \frac{k_{t\;\gamma}}{y x_g s} $. This definition follows closely the definition of the
rescaled transverse momentum $\bar{q}$ in CCFM. In Tab.~\ref{pdfsets} is shown,
that also with this definition a reasonable description of the data can be
achieved, but the parameter $p_0$ of the 
starting distribution is changed significantly.
\begin{table}[htb] 
\begin{center}
\begin{tabular}{| c | c | c | r | r | r | r | }
\hline
\hline
set & $\mu_f $ & $\mu_r  $ & $p_0$ & $k_0$ & $\kt^{cut}$  & $\chi^2/points$  \\
  &   & $ =\sqrt{\pt^2+m^2} $ &   & (GeV)  &  (GeV) & $\chi^2/points$  \\
\hline
 $A0$ &   $\sqrt{\hat{s} + Q_{\prp}^2}$  & $1\cdot\mu_f $ &     0    &   1.33 & 1.33 & 276/248 = 1.1 \\
 $A0-$ &    $\sqrt{\hat{s} + Q_{\prp}^2}$  & $0.5 \cdot \mu_f $  &  -0.01    &   1.33 & 1.33 & 433/248 = 1.75 \\
 $A0+$ &    $\sqrt{\hat{s} + Q_{\prp}^2}$  & $2\cdot \mu_f $  &  -0.01    &   1.33 & 1.33 & 316/248 = 1.3 \\
 $A1$&   $\frac{\pt}{1-z}$    &  &-0.1       &  0.8 & 1.33 &  275/248 = 1.1  \\
\hline
 $B0$ &    $\sqrt{\hat{s} + Q_{\prp}^2}$  & $1\cdot\mu_f $ &     0     &  0.8 & 0.25 & 356/248 = 1.4 \\
 $B0-$  &    $\sqrt{\hat{s} + Q_{\prp}^2}$  & $0.5\cdot\mu_f $ &   0.01    &  0.8 & 0.25 & 483/248 = 1.95 \\
 $B0+$  &    $\sqrt{\hat{s} + Q_{\prp}^2}$  & $2\cdot\mu_f $ &    0.01     &  0.8 & 0.25 & 297/248 = 1.2 \\
 $B1$ &    $\frac{\pt}{1-z}$   & 1 & -0.1        &  0.8 & 0.25 & 702/248 = 2.8  \\
\hline
\hline
GBW &    $\sqrt{\hat{s} + Q_{\prp}^2}$ & 1 & -     &  - & 0.25 & 349/248 = 1.4 \\
\hline
\hline
\end{tabular}
\caption{The different settings of the CCFM un-integrated gluon densities. The initial
distribution is parameterized as  $x {\cal A}_0(x,\kt,Q_0) = N 
 x^{p_0} (1-x)^{p_1} \cdot \exp{\left(-\kt^2/k_0^2\right)}$ with $p_1=4$ and $Q_g=1.33$~GeV. In the last line the
starting distribution from the parameterization of GBW~\protect\cite{GBW} is used.
In the last column, the $\chi^2/points$ to HERA 
$F_2$ data~\protect\cite{H1_F2_1996,H1_F2_2001,ZEUS_F2_1996,ZEUS_F2_2001}
 is given (for $x< 5\cdot 10^{-3}$ and $Q^2 > 4.5 $ GeV$^2$). }
\label{pdfsets}
\end{center}
\end{table}
\section{$\kt$ dependence of \boldmath$x {\cal A}(x,\kt,\mu_f)$}
\begin{figure}[htb]
\vskip -4cm 
\centerline{\rotatebox{0.}{\scalebox{0.6}{\includegraphics{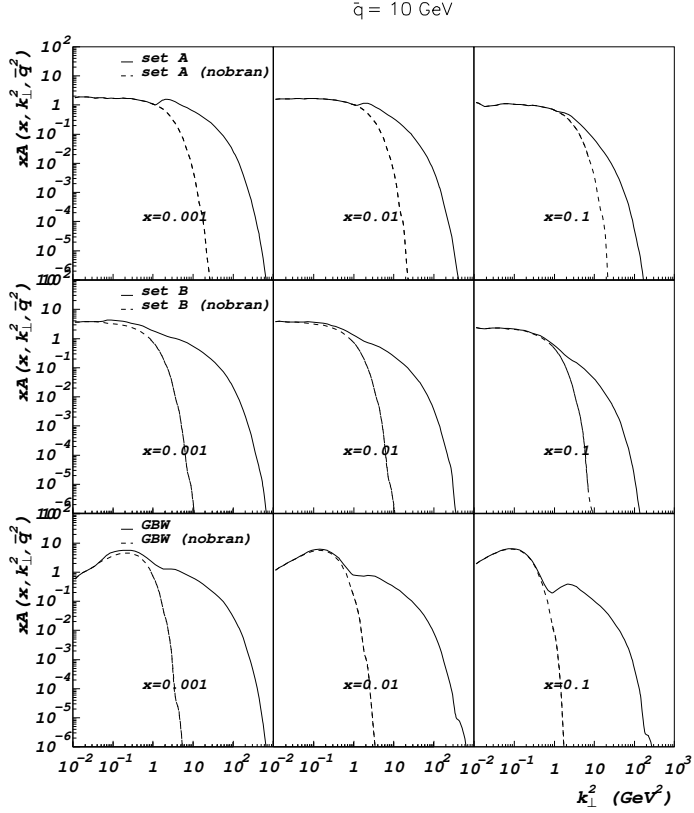}}}}
\vskip -1cm 
 \caption[*]{
 {\it Comparison of the different sets of un-integrated gluon densities obtained
 from the CCFM evolution as described in the text. 
 \label{uPDF}}}
\vskip -1cm 
\end{figure}

In Fig.~\ref{uPDF} the uPDFs (as described in Tab.~\ref{pdfsets}) 
are shown as a function of $\kt$ for different values of $x$. 
Also shown is the contribution of the first term of the rhs of 
eq.(\ref{integral}). It can be seen, that the influence of the intrinsic $\kt$
distribution is concentrated at small values of $\kt$, whereas at values 
$\kt \gap 1 $ GeV the perturbative evolution is responsible. 
The specific choice of the $\kt^{cut}$ results in different distributions at
small values of $\kt$. It is also interesting to observe, that the {\it dip}
 in $\kt$
visible in Fig.~\ref{uPDF} (upper plot) is directly related to $\kt^{cut}$. 
\par
It is interesting to observe, that even with very different starting
distributions, the uPDFs after perturbative evolution are similar (at larger
$\kt$). Thus the small $\kt$ region 
provides new information on the
non-perturbative part of the parton density functions, which can only be
investigated by using uPDFs. Especially the question of saturation effects at
small $\kt$ can be directly investigated with uPDFs. The difference of using a
Gaussian intrinsic $\kt$ distribution to a form motived by the saturation model
of GBW is significant. It is also important to note, that these differences are
washed out for inclusive distributions, as all uPDFs are able to
describe $F_2$ at a similar level. 
\section*{Acknowledgments} 
I am grateful to the organizers of the DIS2004 workshop for the 
stimulating atmosphere and the good organization. 

\end{document}